\documentclass[conference,10pt,letterpaper,final]{IEEEtran}

\usepackage{silence}
\usepackage{cite}

\usepackage{amsmath,amssymb,amsthm,fixmath}

\usepackage{color,colortbl}
\usepackage{xcolor}

\usepackage[inline]{enumitem}

\usepackage{siunitx}
\DeclareSIUnit{\dBm}{dBm}

\usepackage{graphicx}
\usepackage[labelformat=simple]{subcaption}

\usepackage{epstopdf}
\usepackage{cuted}
\usepackage{multirow,booktabs}
\usepackage{diagbox}
\usepackage{makecell}
\usepackage{hhline}
\usepackage{array} 
\newcolumntype{x}{!{\vrule width 2px}}
\newcolumntype{y}{!{\vrule width 1.5px}}

\usepackage{optidef}

\usepackage[hidelinks]{hyperref}

\usepackage[shortcuts,acronym,automake]{glossaries}
\makeglossaries
\newacronym{6g}{6G}{Sixth Generation}
\newacronym{awgn}{AWGN}{additive white Gaussian noise}
\newacronym{bcd}{BCD}{block coordinate descent}
\newacronym{bri}{BRI}{biregular irreducible}
\newacronym{cp}{CP}{control plane}
\newacronym{csi}{CSI}{channel state information}
\newacronym{csit}{CSIT}{channel state information at transmitter}
\newacronym{dft-s-ofdm}{DFT-s-OFDM}{Discrete Fourier Transform-spread-OFDM}
\newacronym{fbl}{FBL}{finite blocklength}
\newacronym{gan}{GAN}{generative adversarial network}
\newacronym{ibl}{IBL}{infinite blocklength}
\newacronym{lfp}{LFP}{leakage-failure probability}
\newacronym{mcs}{MCS}{modulation and coding scheme}
\newacronym{mimo}{MIMO}{multi-input multi-output}
\newacronym{mm}{MM}{Minorize-Maximization}
\newacronym{noma}{NOMA}{non-orthogonal multi-access}
\newacronym{nom}{NOM}{non-orthogonal multiplexing}
\newacronym{ofdm}{OFDM}{orthogonal frequency-division multiplexing}
\newacronym{ofdma}{OFDMA}{orthogonal frequency-division multiple access}
\newacronym{oma}{OMA}{orthogonal multiple access}
\newacronym{papr}{PAPR}{Peak-to-Average Power Ratio}
\newacronym{pdf}{PDF}{probability density function}
\newacronym{per}{PER}{packet error rate}
\newacronym{phy}{PHY}{physical}
\newacronym{pld}{PLD}{physical layer deception}
\newacronym{pls}{PLS}{physical layer security}
\newacronym{prb}{PRB}{physical resource block}
\newacronym{sic}{SIC}{successive interference cancellation}
\newacronym{sinr}{SINR}{signal-to-interference-and-noise ratio}
\newacronym{snr}{SNR}{signal-to-noise ratio}
\newacronym{tdma}{TDMA}{time-division multiple access}
\newacronym{up}{UP}{user plane}
\newacronym{urllc}{URLLC}{ultra-reliable low-latency communication}

\usepackage{tcolorbox}
\tcbuselibrary{many}
\usepackage[norelsize,linesnumbered,ruled]{algorithm2e}
\SetKwRepeat{Do}{do}{while}
\makeatletter
\newcommand{\removelatexerror} {\let\@latex@error\@gobble}
\makeatother

\newcommand{\superscript}[1]{^{\mathrm{#1}}}
\newcommand{\subscript}[1]{_{\mathrm{#1}}}

\usepackage[normalem]{ulem}

\usepackage[textsize=tiny,colorinlistoftodos]{todonotes}

\tikzstyle{note}=[rectangle, minimum width=3cm, draw = none, fill = none, minimum width = 1.5cm, anchor=center, align=left]
\tikzstyle{block}=[rectangle, draw, line width=1pt, fill = none, minimum width = 1cm, minimum height = 0.75cm, anchor=center, inner sep = 0.5mm, align=center]
\tikzstyle{arrow} = [thick,->,>=stealth]

\newif\ifreviewmode
\reviewmodetrue % Set to true to enable review mode
\ifreviewmode
\else
  \renewcommand{\todo}[1]{} % hide todo notes
\fi

\begin{document}

\title{A Semantic Model for Physical Layer Deception}

\author{
	\IEEEauthorblockN{
		Bin~Han\IEEEauthorrefmark{1},
		Yao~Zhu\IEEEauthorrefmark{2},
		Anke~Schmeink\IEEEauthorrefmark{2}, Giuseppe~Caire\IEEEauthorrefmark{3}, and~Hans~D.~Schotten\IEEEauthorrefmark{1}\IEEEauthorrefmark{4}
	}
	\IEEEauthorblockA{
		\IEEEauthorrefmark{1}University of Kaiserslautern (RPTU), \IEEEauthorrefmark{2}RWTH Aachen University,\\  \IEEEauthorrefmark{3}Technical University of Berlin, \IEEEauthorrefmark{4}German Research Center of Artificial Intelligence (DFKI)
	}
}

\bstctlcite{IEEEexample:BSTcontrol}

\maketitle

\begin{abstract}
\Ac{pld} is a novel security mechanism that combines \ac{pls} with deception technologies to actively defend against eavesdroppers. In this paper, we establish a novel semantic model for \ac{pld} that evaluates its performance in terms of semantic distortion. 
By analyzing semantic distortion at varying levels of knowledge on the receiver's part regarding the key, we derive the receiver's optimal decryption strategy, and consequently, the transmitter's optimal deception strategy. 
%Analyzing the semantic distortion under different levels of knowledge about the key at the receiver's end, we derive the optimal decryption strategy of receiver, and therewith the optimal deception strategy of the transmitter. 
The proposed semantic model provides a more generic understanding of the \ac{pld} approach independent from coding or multiplexing schemes, and allows for efficient real-time adaptation to fading channels.
\end{abstract}


\IEEEpeerreviewmaketitle

\glsresetall

\section{Introduction}\label{sec:introduction}
In recent past years, the concept of \ac{pls} has been regaining increasing attention in the field of wireless communications. Without relying on cryptography algorithms, \ac{pls} aims at securing information transmission by exploiting the characteristics of physical channels, and is therefore robust against the emerging threat of quantum computing empowered cyber attacks. 
This is especially crucial in the context of the Internet of Things (IoT) ecosystem, where new services such as ultra-reliable and low-latency communications (URLLC) and massive machine-type communications (mMTC) are becoming prevalent~\cite{PS+2016PLS,CLY+2019PLSURLLC}. These emerging scenarios present unique security challenges that \ac{pls} is well-suited to address.
Consequently, \ac{pls} is widely considered as an important part of the security landscape of future \ac{6g} mobile systems~\cite{PLP+2024security}.

Despite the enhancements in passive security provided by \ac{pls}, wireless security still remains an imbalanced game: it takes barely a risk and cost to attempt eavesdropping, compared to the extensive measures to secure data. To address this issue, we have proposed in \cite{HZS+2023non} and \cite{CHZ+2024physical} a novel framework of \ac{pld}, which combines \ac{pls} and deception technologies to realize an active defense against eavesdroppers by deceiving them with falsified information, while maintaining the ordinary communication over the legitimate channel. By jointly optimizing the encryption coding rate and the power allocation, we managed to simultaneously achieve high secured reliability and effective deception. 

Nevertheless, these preliminary efforts of \ac{pld} are still limited in several aspects. First, they are specified to power-domain non-orthogonal multiplexing schemes, which may limit the integration of \ac{pld} with conventional wireless technologies. Second, the optimization problem is set up regarding leakage-failure probability and effective deception, which may lack generality in different applications. Third, the joint optimization of the encryption coding rate and the power allocation can be computationally expensive for real-time adaptation to fading channels.

In this work, we aim at addressing these limitations by proposing a semantic model for \ac{pld}, which invokes a more generic metric, i.e., the semantic distortion, for evaluating communication and security performance. The semantic model is more general and flexible regarding system implementation, as it relies on no specific coding or multiplexing scheme. Moreover, by analyzing the semantic distortion in \ac{pld} systems, we are able to derive the optimal decryption strategy with respect to channel conditions, which applies for both the legitimate receiver and the eavesdropper. This further allows us to derive the optimal deception strategy of the transmitter, which can be efficiently implemented by adjusting the deceiving probability upon measured/estimated channel conditions in real-time, without the need of solving complex non-linear optimization problems.

The remainder of this paper is organized as follows. First, in Sec.~\ref{sec:semcom_model} we introduce the system model for \ac{pld},  establish its semantic model consisting of two transport channels, and derive the transport channel models. 
We then proceed with Sec.~\ref{sec:distortion} to analyze the semantic distortion at varying levels of knowledge about the key on the receiver’s side. 
%Then we follow with Sec.~\ref{sec:distortion} to analyze the semantic distortion, upon different levels of knowledge about the key at the receiver's end. 
In Sec.~\ref{sec:opportunistic_rx}, we propose and analyze three decryption strategies for cases where the receiver is unable to obtain a valid key from the received signal.
%Therewith, in Sec.~\ref{sec:opportunistic_rx} we propose and analyze three different decryption strategies, when the receiver fails to obtain any valid key from the received signal. 
The optimal opportunistic selection among these strategies is derived as a simple linar programming problem. Subsequently, the optimal deception strategy of transmitter is discussed in Sec.~\ref{sec:tx_strategy_optimization}. Finally, we present numerical results in Sec.~\ref{sec:numerical_results} and conclude the paper in Sec.~\ref{sec:conclusion}.

\section{Models}\label{sec:semcom_model}
\subsection{System Model}
\begin{figure}[!htbp]
	\centering
	\includegraphics[width=0.9\linewidth]{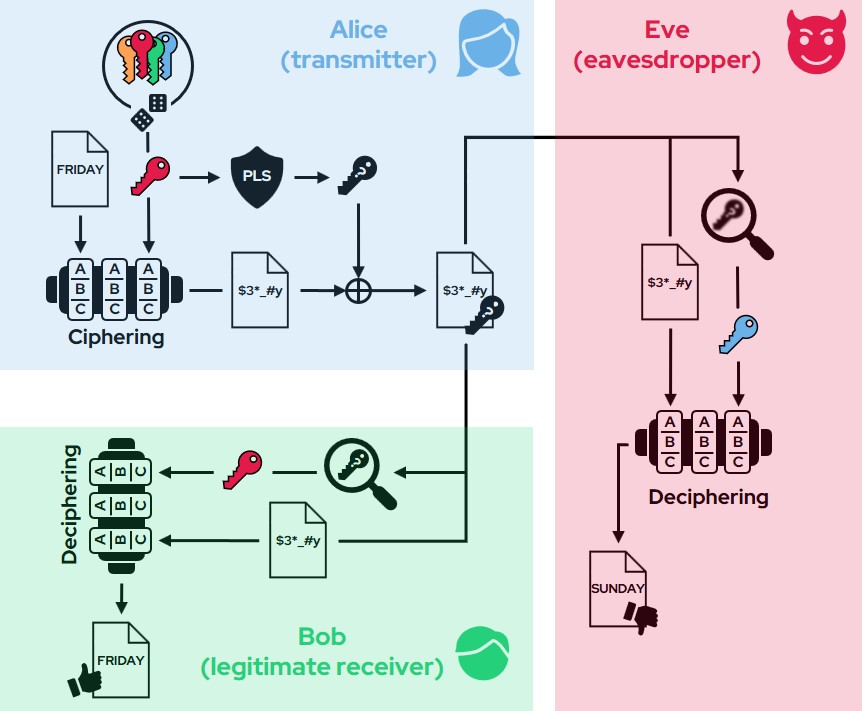}
	\caption{System model for physical layer deception}
	\label{fig:sys_model}
\end{figure}

The principle of \ac{pld} is illustrated in Fig.~\ref{fig:sys_model}. The transmitter, \emph{Alice}, ciphers every plaintext message with an individually selected random key. The key, which is not known by anyone but \emph{Alice} a priori, is then multiplexed with its associated ciphertext for a joint transmission. Unlike conventional \ac{pls} approaches that attempt to secure the entire transmitted message from eavesdropping, in \ac{pld} scenarios, \ac{pls} is selectively applied on the key part, leaving the ciphertext well exposed to the eavesdropper. With a carefully designed crypto, combining the correct ciphertext with an incorrect key, which are likely to be obtained by the eavesdropper, \emph{Eve}, will generate a falsified message, which will deceive \emph{Eve}. Meanwhile, benefiting from the superior channel gain, the legitimate user, \emph{Bob}, is supposed to obtain both the ciphertext and the associated key, and therefore capable of recovering the original data.
\vspace{-3mm}

It shall be noted that when \emph{Eve} is aware of the \ac{pld} mechanism, it may exploit it even without correctly decoding the key. For instance, considering its own deciphering result likely to be a falsified message, \emph{Eve} may exclude it from the possible candidates of the original plaintext, and thus gain partial information about the plaintext. To encounter such eavesdroppers with deep insights about the \ac{pld} mechanism, we proposed in \cite{CHZ+2024physical} an advanced deception strategy where
\begin{enumerate*}[label=\emph{\roman*)}]
	\item every valid ciphertext codeword must also be a valid plaintext codeword,
	\item the chance of confusing a codeword (either ciphertext or key) with another valid codeword is negligible compared to that of decoding failure, and
	\item the deceptive ciphering can be randomly activated/deactivated. When deactivated, no key is generated, the plaintext is unciphered and multiplexed with a litter sequence instead.
\end{enumerate*}
Here, the litter sequence is no valid key but to prevent \emph{Eve} from estimating the activation status by observing the power profile of the transmitted signal.

\subsection{Semantic Perspective}
\begin{figure}[!htbp]
	\centering
	\scalebox{0.8}{
	\begin{tikzpicture}[node distance = 1.6cm, auto]		
		\node(stx)[note]{$w\in\mathbb{S}$};
		\node(semenc)[block, right of = stx]{$u_k$};
		\node(chenc)[block, right of = semenc]{$\phi\superscript{p}$};
		\node(legtch1)[block, right of = chenc, draw=none]{};
		\node(prmch)[block, right of = chenc, minimum width = 5cm, minimum height = 1.5cm, yshift=2mm, dashed]{};
		\node()[right of = chenc, yshift=7mm]{$c\superscript{p}$};
		\node(chdec)[block, right of = legtch1]{$\eta\superscript{p}$};
		\node(semdec)[block, right of = chdec]{$v_{\hat k}$};
		\node(srx)[note, right of = semdec]{$\hat w\in\mathbb{S}^+$};
		\draw[arrow](stx)--(semenc);
		\draw[arrow](semenc)--(chenc) node[midway,above]{$s$};
		\draw[arrow](chenc)--(legtch1);
		\draw[arrow](legtch1)--(chdec);
		\draw[arrow](chdec)--(semdec) node[midway,above]{$\hat s$};
		\draw[arrow](semdec)--(srx);

		\node(ktx)[note,below of = stx]{$k\in\mathbb{K}^+$};
		\node(x)[inner sep = 0, minimum width = 0] at (ktx-|semenc){};
		\draw[arrow](ktx)--(x)--(semenc);
		\node(chenc2)[block, below of = chenc]{$\phi\superscript{s}$};
		\node(legtch2)[block, below of = legtch1, draw=none]{};
		\node(scdch)[block, right of = chenc2, minimum width = 5cm, minimum height = 1.5cm, yshift=-2mm, dashed]{};
		\node()[right of = chenc2, yshift=-7mm]{$c\superscript{s}$};
		\node(chdec2)[block, below of = chdec]{$\eta\superscript{s}$};
		\node(x2)[inner sep = 0, minimum width = 0] at (ktx-|semdec){};
		\draw[arrow](x)--(chenc2);
		\draw[arrow](chenc2)--(legtch2);
		\draw[arrow](legtch2)--(chdec2);
		\draw[arrow](chdec2)--(x2)--(semdec) node[midway,right]{$\hat k\in\mathbb{K}^+$};

		\node(legtch)[block, below of = legtch1, minimum height = 2.5cm, yshift = 0.9cm]{$T$};
	\end{tikzpicture}
	}
	\caption{Dual-channel model of \ac{pld}}
	\label{fig:sem_ch_model}
\end{figure}
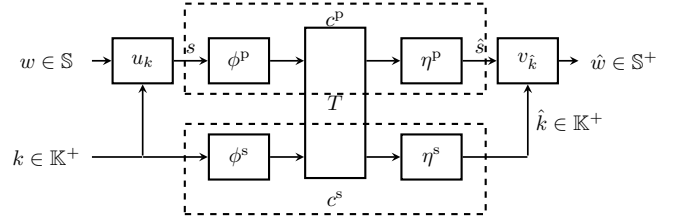

From a semantic communication perspective, the overall model of a communication system deploying the \ac{pld} mechanism is illustrated in Fig.~\ref{fig:sem_ch_model}. Every plaintext to be sent can be considered as a meaning $w\in\mathbb{S}$. For each meaning, \emph{Alice} encodes it into a message(ciphertext) $s\in\mathbb{S}$ with an semantic encoder (encryptor) $u_k$, where $k\in\mathbb{K}$ is a randomly selected key, and $\mathbb{S}$ and $\mathbb{K}$ are the sets of all feasible plaintext codewords and keys, respectively. Alternatively, with deception deactivated, \emph{Alice} sends the original plaintext unciphered, i.e. $s=w$, together with a litter sequence, which is denoted as $k\subscript{NULL}$. Thus, by defining $\mathbb{K}^+=\mathbb{K}\cup\{k\subscript{NULL}\}$, we can unify both cases into a generic model, where every $w$ is always semantically encoded by $u_k$ with $k\in\mathbb{K}^+$, and particularly $u_{k\subscript{NULL}}(w)=w$ for all $w\in\mathbb{S}$.

The message $s$ is then transmitted over the primary transport channel $c\superscript{p}$, which consists of the primary channel encoder $\phi\superscript{p}$, the physical channel $T$, and the primary channel decoder $\eta\superscript{p}$. Similarly, the key $k$ is transmitted over the secondary transport channel $c\superscript{s}$ that consists of the secondary channel encoder $\phi\superscript{s}$, the physical channel $T$ and the secondary channel decoder~$\eta\superscript{s}$. \emph{Bob} receives both the message $\hat s\in\mathbb{S}^+$ over the primary transport channel and the key $\hat k\in\mathbb{K}^+$ over the secondary transport channel, where $\mathbb{S}^+=\mathbb{S}\cup\{s\subscript{NULL}\}$ and $s\subscript{NULL}$ stands for the decoding error flag. Subsequently, \emph{Bob} semantically decodes $\hat s$ with $\hat k$ to estimate the meaning $\hat w\in\mathbb{S}^+$.
Note the similarity between this dual-classical-channel model and the classical-quantum semantic communication model discussed in \cite{BCD+2022semantic}
and \cite{BCW2022mosaics}%
, with our encryptor $u_k$ behaving as a \ac{bri} function and $k$ as the the random seed.

In this work, we consider a worst case where \emph{Eve} shares the same level of knowledge about the system as \emph{Bob}, and is as well able to intercept the message $\hat s$ and the key $\hat k$ over the primary and secondary transport channels, respectively. The only difference between \emph{Eve} and \emph{Bob} is that the former has a worse physical channel condition, which can be practically achieved with accurate beamforming. Thus, the wiretap channel shall be identically modeled as the legitimate channel. In practice, the potentially existing eavesdroppers usually have less knowledge about the system specifications than the legitimate receiver, which can only worsen the eavesdropping performance, and thus is not discussed in this paper.
\vspace{-3mm}

\vspace{-8mm}
\subsection{Transport Channel Model}\label{sec:tr_ch_model}
\begin{figure}[!htbp]
	\centering
	\scalebox{0.9}{
	\begin{tikzpicture}[node distance = 1.5cm, auto]		
		\node(stx)[note]{$s$};
		\node(srx)[note, right of=stx, xshift=2cm, yshift = 0.5cm]{$s$};
		\node(snull)[note, right of=stx, xshift=2cm, yshift = -0.5cm]{? ($s\subscript{NULL}$)};
        \node(note)[note, right of=snull, xshift=1cm, yshift = 0.5cm]{$\forall s\in\mathbb{S}$};
		\draw[arrow](stx)--node[midway,above]{$1-\varepsilon\superscript{p}$}(srx);
		\draw[arrow](stx)--node[midway,below]{$\varepsilon\superscript{p}$}(snull);
	\end{tikzpicture}
	}
	\caption{Model of the primary transport channel}
	\label{fig:prim_phy_channel}
\end{figure}
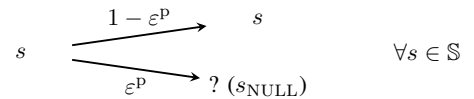

The primary transport channel, under proper channel coding and error detection, is an erasure channel. The primary transport channel has therefore the conditional \ac{pdf}
\begin{equation}
	\begin{split}
	c\superscript{p}(\hat s\vert s)=&\begin{cases}
		1-\varepsilon\superscript{p}, & \hat s=s\\
		\varepsilon\superscript{p}, & \hat s=s\subscript{NULL}\\
		0, & \text{otherwise}
	\end{cases}\\
	=&(1-\varepsilon\superscript{p})\delta(\hat s-s)+\varepsilon\superscript{p}\delta(\hat s-s\subscript{NULL}),
	\end{split}
	\label{eq:prim_phy_channel}
\end{equation}
and can be illustrated by Fig.~\ref{fig:prim_phy_channel}.
\vspace{-2mm}

Due to the possible transmission of a litter sequence instead of a valid key, the secondary transport channel%
, with sufficient channel coding redundancy and error detection, makes
% is not a simple erasure channel but 
a Z-channel like illustrated in Fig.~\ref{fig:sec_phy_channel}:
\begin{equation}
	\begin{split}
		c\superscript{s}(\hat k\vert k)=&\begin{cases}
			1-\varepsilon\superscript{s}, & \hat k=k, k\in\mathbb{K}\\
			\varepsilon\superscript{s}, & \hat k=k\subscript{NULL}, k\in\mathbb{K}\\
			1, & \hat k=k=k\subscript{NULL}\\
			0, & \text{otherwise}
		\end{cases}\\
		=&\left[(1-\varepsilon\superscript{s})\delta(\hat k -k)+\varepsilon\superscript{s}\delta(\hat k-k\subscript{NULL})\right]\cdot\\
		&\left[1-\delta(k-k\subscript{NULL})\right]+\delta(\hat k-k\subscript{NULL})
	\end{split}
	\label{eq:sec_phy_channel}
\end{equation}
\vspace{-2mm}
\begin{figure}[!htbp]
	\centering
	\scalebox{0.9}{
	\begin{tikzpicture}[node distance = 1.5cm, auto]		
		\node(knulltx)[note]{$k\subscript{NULL}$};
		\node(ktx)[note,below of=knulltx]{$k$};
		\node(krx)[note, right of=ktx, xshift=2cm]{$k$};
		\node(knullrx)[note, right of=knulltx, xshift=2cm,]{? ($k\subscript{NULL}$)};
        \node(note)[note, right of=krx, xshift=1cm, yshift = 0.75cm]{$\forall k\in\mathbb{K}$};
		\draw[arrow](ktx)--node[midway,below]{$1-\varepsilon\superscript{s}$}(krx);
		\draw[arrow](ktx)--node[midway,above]{$\varepsilon\superscript{s}$}(knullrx);
		\draw[arrow](knulltx)--node[midway,above]{$1$}(knullrx);
	\end{tikzpicture}
	}
	\caption{Model of the secondary transport channel}
	\label{fig:sec_phy_channel}
\end{figure}
\vspace{-2mm}

\section{Distortion Analysis}\label{sec:distortion}
\subsection{Semantic Distortion with Synchronized Key}
Given a certain key $k$ that is shared on both sides of the channel, the encryptor-decryptor pair of the \ac{pld} scheme can be considered as a pair of semantic encoder/decoder, which according to \cite{SCG2024theory} can be stochastically characterized by the conditional \ac{pdf} of their output upon input, i.e., $u_k(s\vert w)$ and $v_k(\hat w\vert \hat s)$, respectively. The corresponding semantic channel can be then characterized by
\begin{equation}
	\psi_k(\hat w\vert w)=\sum\limits_{s,\hat s}u_k(s\vert w)c\superscript{p}(\hat s\vert s)v_k(\hat w\vert \hat s),
	\label{eq:semantic_channel_key_synchronized}
\end{equation}
and the average ciphertext distortion achieved thereover is
\begin{equation}
	\begin{split}
		D_k=&\sum\limits_{w,\hat w}p(w)\psi_k\superscript{p}(\hat w\vert w)d(w,\hat w)\\
		=&\sum\limits_{w,s,\hat s,\hat w}p(w)u_k(s\vert w)c\superscript{p}(\hat s \vert s)v_k(\hat w \vert \hat s)d(w,\hat w),
	\end{split}
	\label{eq:distortion_key_synchronized}	
\end{equation}
where $d(w_1,w_2)$ is the semantic distortion between the pair of meanings $[w_1,w_2]\in\left(\mathbb{S}^+\right)^2$. Generally, it holds  
\begin{equation}
	d(w,w)=0,\quad w\in\mathbb{S}^+.
	\label{eq:zero_semantic_distance}	
\end{equation}

\subsection{Semantic Distortion with Inaccurate Key}
Now, we consider that the receiver does not have the key $k$ synchronized from the transmitter with guaranteed accuracy, but takes an estimate $\hat k$ instead. This has no impact on the semantic encoder but on the decoder, which becomes $v_{\hat k}(\hat w\vert\hat s)$. Given a certain pair $[k,\hat k]$, the semantic channel in Eq.~\eqref{eq:semantic_channel_key_synchronized} and its average ciphertext distortion in Eq.~\eqref{eq:distortion_key_synchronized} shall be revised into
\begin{equation}
	\psi_{k,\hat k}(\hat w\vert w)=\sum\limits_{s,\hat s}u_k(s\vert w)c\superscript{p}(\hat s\vert s)v_{\hat k}(\hat w\vert \hat s), 
	\label{eq:semantic_channel_key_estimated}
\end{equation}
% and
\begin{equation}
	\begin{split}
	D_{k,\hat k}=&\sum\limits_{w,\hat w}p(w)\psi_{k,\hat k}(\hat w\vert w)d(w,\hat w)\\
	=&\sum\limits_{w,s,\hat s,\hat w}p(w)u_k(s\vert w)c\superscript{p}(\hat s \vert s)v_{\hat k}(\hat w \vert \hat s)d(w,\hat w).
	\end{split}
	\label{eq:distortion_key_estimated}
\end{equation}
% respectively.

\subsection{Semantic Distortion with Transmitted Key}
Now, consider random key $k\in\mathbb{K}^+$ where $\mathbb{K}^+=\mathbb{K}\cup\{k\subscript{NULL}\}$, which is transmitted over the secondary transport channel $c\superscript{s}(\hat k \vert k).$ We have
\begin{equation}
	p(k,\hat k)=p(k)c\superscript{s}(\hat k\vert k).
\end{equation}

Thus, the overall semantic channel with dual transport channels becomes
\begin{equation}
	\begin{split}
		\psi(\hat w\vert w)=&\sum\limits_{k,\hat k}p(k,\hat k)\psi_{k,\hat k}(\hat w\vert w)\\
		=&\sum\limits_{s,\hat s,k,\hat k}p(k)u_k(s\vert w)c\superscript{p}(\hat s\vert s)c\superscript{s}(\hat k\vert k)v_k(\hat w\vert \hat s),	
	\end{split}
	\label{eq:semantic_channel_key_transmitted}
\end{equation}
and its average semantic distortion is
\begin{align}
	D=&\sum\limits_{k,\hat k}p(k,\hat k)D_{k,\hat k}\nonumber\\
	% =&\sum\limits_{\substack{k,w,s,\\\hat s,\hat w,\hat k}}p(k,\hat k)p(w)u_k(s\vert w)c\superscript{p}(\hat s\vert s)v_{\hat k}(\hat w\vert \hat s)d(w,\hat w)\nonumber\\
	=&\sum\limits_{\substack{k,w,s,\\\hat s,\hat w,\hat k}}p(k)p(w)u_k(s\vert w)c\superscript{p}(\hat s\vert s)c\superscript{s}(\hat k\vert k)v_{\hat k}(\hat w\vert \hat s)d(w,\hat w)\nonumber\\
	=&\sum\limits_{\substack{w,\hat w}}p(w)\psi(\hat w\vert w)d(w,\hat w), 
	\label{eq:distortion_key_transmitted}
\end{align}
which is achieved by \emph{Bob} who possesses full knowledge about the mechanism and system specifications.

\subsection{Semantic Distortion with Deterministic Cryptosystem}
So far we have followed the semantic communication model to interpret the encryptor and decryptor as stochastic functions. It shall be remarked, however, given the key $k$, the encryptor and decryptor can also be deterministically characterized as $s=f_k(w)$ and $\hat w=f^{-1}_k(\hat s)$, respectivelly. Thus, we have:
\begin{align}
	u_k(s\vert w)&=\begin{cases}
		1, & s=f_k(w),\\
		0, & \text{otherwise},
	\end{cases}\label{eq:pdf_encryptor}\\
	v_{\hat k}(\hat w\vert \hat s)&=\begin{cases}
		1, & \hat w=f^{-1}_{\hat k}(\hat s),\\
		0, & \text{otherwise}, 
	\end{cases}\label{eq:pdf_decryptor}
\end{align}

\begin{figure*}[!htbp]
	\setcounter{equation}{\theequation+9}
	\begin{align}
		% \begin{split}
		D=&\sum\limits_{w, \hat k}p(w)\left\{\sum\limits_{k\in\mathbb{K}}p(k)c\superscript{s}(\hat k\vert k)\left[(1-\varepsilon\superscript{p})d\left(w,f^{-1}_{\hat k}\left(f_k(w)\right)\right)+\varepsilon\superscript{p}d\left(w,s\subscript{NULL}\right)\right]\right.\nonumber\\
		&\left.+(1-\alpha)c\superscript{s}(\hat k\vert k\subscript{NULL})\left[(1-\varepsilon\superscript{p})d\left(w,f^{-1}_{\hat k}\left(f_{k\subscript{NULL}}(w)\right)\right)+\varepsilon\superscript{p}d\left(w,s\subscript{NULL}\right)\right]\right\}\nonumber\\
		%
		% \overset{\eqref{eq:enc_null_key}}{=}&\left.\sum\limits_{w, \hat k}p(w)\right\{\sum\limits_{k\in\mathbb{K}}p(k)c\superscript{s}(\hat k\vert k)\left[(1-\varepsilon\superscript{p})d\left(w,f^{-1}_{\hat k}\left(f_k(w)\right)\right)+\varepsilon\superscript{p}d\left(w,s\subscript{NULL}\right)\right]\nonumber\\
		% &\left.+(1-\alpha)c\superscript{s}(\hat k\vert k\subscript{NULL})\left[(1-\varepsilon\superscript{p})d\left(w,f^{-1}_{\hat k}\left(w\right)\right)+\varepsilon\superscript{p}d\left(w,s\subscript{NULL}\right)\right]\right\}\nonumber\\
		% %
		% \overset{\eqref{eq:sec_phy_channel}}{=}&\sum\limits_{w, \hat k}p(w)\left\{\sum\limits_{k\in\mathbb{K}}p(k)\left[\left(1-\varepsilon\superscript{s}\right)\left((1-\varepsilon\superscript{p})d\left(w,f^{-1}_{k}\left(f_k(w)\right)\right)+\varepsilon\superscript{p}d\left(w,s\subscript{NULL}\right)\right)\right.\right.\nonumber\\
		% &\left.\left.+\varepsilon\superscript{s}\left((1-\varepsilon\superscript{p})d\left(w,f^{-1}_{k\subscript{NULL}}\left(f_k(w)\right)\right)+\varepsilon\superscript{p}d\left(w,s\subscript{NULL}\right)\right)\right]+(1-\alpha)\left[(1-\varepsilon\superscript{p})d\left(w,f^{-1}_{k\subscript{NULL}}\left(w\right)\right)+\varepsilon\superscript{p}d\left(w,s\subscript{NULL}\right)\right]\right\}\nonumber\\
		% %
		\underset{\eqref{eq:sec_phy_channel}}{\overset{\eqref{eq:enc_null_key}}{=}}&\sum\limits_{w, \hat k}p(w)\left\{\sum\limits_{k\in\mathbb{K}}p(k)\left[\left(1-\varepsilon\superscript{s}\right)\left((1-\varepsilon\superscript{p})d\left(w,f^{-1}_{k}\left(f_k(w)\right)\right)+\varepsilon\superscript{p}d\left(w,s\subscript{NULL}\right)\right)\right.\right.\nonumber\\
		&\left.\left.+\varepsilon\superscript{s}\left((1-\varepsilon\superscript{p})d\left(w,f^{-1}_{k\subscript{NULL}}\left(f_k(w)\right)\right)+\varepsilon\superscript{p}d\left(w,s\subscript{NULL}\right)\right)\right]+(1-\alpha)\left[(1-\varepsilon\superscript{p})d\left(w,f^{-1}_{k\subscript{NULL}}\left(w\right)\right)+\varepsilon\superscript{p}d\left(w,s\subscript{NULL}\right)\right]\right\}\nonumber\\
		\underset{\eqref{eq:correct_enc_dec}}{\overset{\eqref{eq:dec_null_key}}{=}}&\sum\limits_{w, \hat k}p(w)\left\{\sum\limits_{k\in\mathbb{K}}p(k)\left[\left(1-\varepsilon\superscript{s}\right)\left((1-\varepsilon\superscript{p})d(w,w)+\varepsilon\superscript{p}d\left(w,s\subscript{NULL}\right)\right)\right.\right.\label{eq:distortion_with_random_activation}\\
		&\left.\left.+\varepsilon\superscript{s}\left((1-\varepsilon\superscript{p})d\left(w,f_k(w)\right)+\varepsilon\superscript{p}d\left(w,s\subscript{NULL}\right)\right)\right]+(1-\alpha)\left[(1-\varepsilon\superscript{p})d(w,w)+\varepsilon\superscript{p}d\left(w,s\subscript{NULL}\right)\right]\right\}\nonumber\\
		\overset{\eqref{eq:zero_semantic_distance}}{=}&\sum\limits_{w, \hat k}p(w)\left\{\sum\limits_{k\in\mathbb{K}}p(k)\left[\left(1-\varepsilon\superscript{s}\right)\varepsilon\superscript{p}d\left(w,s\subscript{NULL}\right)+\varepsilon\superscript{s}\left((1-\varepsilon\superscript{p})d\left(w,f_k(w)\right)+\varepsilon\superscript{p}d\left(w,s\subscript{NULL}\right)\right)\right]+(1-\alpha)\varepsilon\superscript{p}d\left(w,s\subscript{NULL}\right)\right\}\nonumber\\
		%
		% =&\sum\limits_{w, \hat k}p(w)\left\{\sum\limits_{k\in\mathbb{K}}p(k)\left[\varepsilon\superscript{p}d\left(w,s\subscript{NULL}\right)+\varepsilon\superscript{s}(1-\varepsilon\superscript{p})d\left(w,f_k(w)\right)\right]+(1-\alpha)\varepsilon\superscript{p}d\left(w,s\subscript{NULL}\right)\right\}\nonumber\\
		% %
		=&\underset{\text{Distortion from loss}}{\underbrace{\varepsilon\superscript{p}d\left(w,s\subscript{NULL}\right)}}+\underset{\text{Distortion from confusion (deception)}}{\underbrace{\left(1-\varepsilon\superscript{p}\right)\varepsilon\superscript{s}\sum\limits_{w}p(w)\sum\limits_{k\in\mathbb{K}}p(k)d\left(w,f_k(w)\right)}}\nonumber
		% \end{split}
	\end{align}
	\hrule
\end{figure*}

\setcounter{equation}{\theequation-10}
which assemble into
\begin{equation}
	u_k(s\vert w)v_{\hat k}(\hat w\vert \hat s)=\begin{cases}
		1, & [s,\hat w] = \left[f_k(w),f^{-1}_{\hat k}(\hat s)\right]\\
		0, & \text{otherwise}	
	\end{cases}.
	\label{eq:deterministic_integrated_enc_dec}
\end{equation}

Particularly, we have
\begin{align}
	f_{k\subscript{NULL}}(w)=w,\quad &\forall w\in\mathbb{S}\label{eq:enc_null_key},\\
	f^{-1}_{k\subscript{NULL}}(\hat s)=\hat s,\quad &\forall \hat s\in\mathbb{S}^+\label{eq:dec_null_key},\\
	f^{-1}_{\hat k}(s\subscript{NULL})=s\subscript{NULL},\quad &\forall \hat k\in\mathbb{K}^+,\label{eq:dec_null_msg}\\
	f^{-1}_{k}(f_k(w))=w,\quad &\forall [w,k]\in\mathbb{S}\times\mathbb{K}^+.\label{eq:correct_enc_dec}
\end{align}

With Eq.~\eqref{eq:deterministic_integrated_enc_dec}, we can rewrite Eqs.~\eqref{eq:distortion_key_synchronized}, \eqref{eq:distortion_key_estimated}, and \eqref{eq:distortion_key_transmitted} as
\begin{align}
	&D_k=\sum\limits_{w,\hat s}p(w)c\superscript{p}\left(\hat s\vert f_k(w)\right)d\left(w,f^{-1}_k(\hat s)\right),
	\label{eq:distortion_key_synchronized_alt}\\
	&D_{k,\hat k}=\sum\limits_{w,\hat s}p(w)c\superscript{p}\left(\hat s\vert f_k(w)\right)d\left(w,f^{-1}_{\hat k}(\hat s)\right),
	\label{eq:distortion_key_estimated_alt}
\end{align}
% and
\begin{equation}\hspace{-4mm}
D=\sum\limits_{k,w,\hat s,\hat k}p(k)p(w)c\superscript{p}\left(\hat s\vert f_k(w)\right)c\superscript{s}(\hat k\vert k)d\left( w,f^{-1}_{\hat k}(\hat s)\right).
\label{eq:distortion_overall}
\end{equation}
% respectively.

Now, we take into account the primary transport channel model~\eqref{eq:prim_phy_channel} and the primary decode failure case~\eqref{eq:dec_null_msg}, applying them to Eq.~\eqref{eq:distortion_overall} to obtain
\begin{equation}
	\begin{split}
	D\overset{\eqref{eq:prim_phy_channel}}{=}&\sum\limits_{k,w,\hat k}p(k)p(w)c\superscript{s}(\hat k\vert k)
	[(1-\varepsilon\superscript{p})\\
	&\left.d\left(w,f^{-1}_{\hat k}\left(f_k(w)\right)\right)+\varepsilon\superscript{p}d\left(w,f^{-1}_{\hat k}(s\subscript{NULL})\right)\right]\\
	\overset{\eqref{eq:dec_null_msg}}{=}&\sum\limits_{k,w,\hat k}p(k)p(w)c\superscript{s}(\hat k\vert k)
	[(1-\varepsilon\superscript{p})\\
	&\left.d\left(w,f^{-1}_{\hat k}\left(f_k(w)\right)\right)+\varepsilon\superscript{p}d\left(w,s\subscript{NULL}\right)\right].
	\end{split}
	\label{eq:distortion_overall_with_prim_phy_channel}
\end{equation}

Furthermore, we consider the deceptive encryptor to be randomly activated by the transmitter with chance $\alpha\in[0,1]$, where the key $k$ is randomly drawn from the set $\mathbb{K}$, i.e., $\sum\limits_{k\neq k\subscript{NULL}}p(k)=\alpha$. The chance of deactivation is then $1-\alpha$ where no key is generated, i.e., $p(k\subscript{NULL})=1-\alpha$. Thus, from Eq.~\eqref{eq:distortion_overall_with_prim_phy_channel} we can derive Eq.~\eqref{eq:distortion_with_random_activation}.

\begin{figure*}[!htbp]
\setcounter{equation}{\theequation+8}
\begin{equation}
    \begin{split}
        \tilde D=&\varepsilon\superscript{p}D\subscript{loss}+(1-\varepsilon\superscript{p})\sum\limits_{w, \hat w}p(w)\left[\varepsilon\superscript{s}\sum\limits_{k\in\mathbb{K}}p(k)\tilde v_{k\subscript{NULL}}\left(\hat w\vert f_k(w)\right)d\left(w,\hat w\right)+(1-\alpha)\tilde v_{k\subscript{NULL}}\left(\hat w\vert w\right)d\left(w,\hat w\right)\right]\\
        =&\varepsilon\superscript{p}D\subscript{loss}+(1-\varepsilon\superscript{p})\left\{\beta_1\underset{\Delta_1}{\underbrace{\varepsilon\superscript{s}\alpha D\subscript{conf}}}+\beta_2\underset{\Delta_2}{\underbrace{\left[\varepsilon\superscript{s}\alpha+(1-\alpha)\right]D\subscript{loss}}}+\beta_3\underset{\Delta_3}{\underbrace{\left[\varepsilon\superscript{s}\frac{\alpha(\mathcal{S}-2)}{\mathcal{S}-1}D\subscript{conf}+(1-\alpha)D\subscript{conf}\right]}}\right\}
    \end{split}\label{eq:distortion_opportunistic_decryptor}
\end{equation}
\hrule
\end{figure*}
\setcounter{equation}{\theequation-8}

Without loss of generality, when considering for all $[k,w,w']\in\mathbb{K}\times\mathbb{S}^2$ that
\begin{align}
	f_k(w)&\neq w,\label{eq:effective_enc}\\
	d(w,f_k(w))&=d(w', f_k(w'))=D\subscript{conf},\\
	d(w,s\subscript{NULL})&=D\subscript{loss},
\end{align}
Eq.~\eqref{eq:distortion_with_random_activation} can be further simplified into
\begin{equation}
	D=\varepsilon\superscript{p}D\subscript{loss}+\alpha\left(1-\varepsilon\superscript{p}\right)\varepsilon\superscript{s}D\subscript{conf}.
	\label{eq:distortion_simplified}
\end{equation}
In scenarios of \ac{pld}, it commonly holds $D\subscript{conf}> D\subscript{loss}>0$.
% Taking its partial derivative with respect to $\varepsilon\superscript{p}$, we have
% \begin{equation}
% 	\begin{split}
% 	\frac{\partial D}{\partial \varepsilon\superscript{p}}=D\subscript{loss}-\left[\varepsilon\superscript{s}-(1-\varepsilon\superscript{p})\frac{\partial\varepsilon\superscript{s}}{\partial\varepsilon\superscript{p}}\right]D\subscript{conf}.	
% 	\end{split}
% 	\label{eq:partial_derivative}
% \end{equation}
% In scenarios of \ac{pld}, it commonly holds $D\subscript{conf}> D\subscript{loss}>0$.

% \section{Secrecy and Deception Analyses}\label{sec:analyses}
% Now consider a classical wiretap channel model where \emph{Alice} tries to transmit information to \emph{Bob}, while \emph{Eve} tries to eavesdrop the communication. Without loss of generality we consider that \emph{Bob} has superiorer \ac{sinr} than \emph{Eve} for both the primary and secondary transport channels, i.e. $\gamma\bob\superscript{p}>\gamma\eve\superscript{p}$ and $\gamma\bob\superscript{s}>\gamma\eve\superscript{s}$.

\section{Opportunistic Decryptor}\label{sec:opportunistic_rx}
Obviously, the semantic distortion from confusion $\alpha\left(1-\varepsilon\superscript{p}\right)\varepsilon\superscript{s}D\subscript{conf}$ plays a critical role in the overall semantic distortion $D$. Especially, when $\varepsilon\superscript{p}\ll 1$ and $D\subscript{conf}\gg D\subscript{loss}$, it dominates the overall distortion. Hence, we seek for alternative strategies to reduce the confusion distortion by leveraging knowledge about the \ac{pld} mechanism design and the system specifications.
We consider three options for the receiver when obtaining a $\hat k=k\subscript{NULL}$ over the secondary transport channel:
\begin{enumerate}
	\item \emph{Perception}: The receiver assumes that no key was decoded because no key was transmitted, i.e., $\hat k=k\subscript{NULL}$. It follows \eqref{eq:dec_null_key} to perceive the decoded as an unencrypted plaintext message, i.e., $\hat w=\hat s$. This is also the default option that is applied by deterministic decryptors.
	\item \emph{Dropping}: Unable to decide if the key was not transmitted or only not decoded, the receiver simply drops the received message $\hat s$ without further consideration, i.e., $\hat w=s\subscript{NULL}$.
	\item \emph{Exclusion}: The receiver assumes that a valid key was used and transmitted, but not correctly decoded. It excludes therewith the possibility of $w=\hat {s}$ according to \eqref{eq:effective_enc}, and thus selects another codeword $\hat w\in\mathbb{S}$ that $\hat w\neq \hat s$. In this case, the distortion-optimal strategy to select $\hat w$ shall follow the principle of maximal likelihood, i.e.,
	\begin{equation}
		f^{-1}_{\hat k}(\hat s)=\arg\max\limits_{w\in\mathbb{S}\backslash\{\hat s\}}p(w).\label{eq:exclusion_strategy_generic}
	\end{equation}
	
In case of multiple optimal solutions, an arbitrary selection is equivalent. The proof is trivial and omitted here. Note that this assumes no specific distribution of $p(w)$. For convenience of discussion, we consider in the remainder of this paper $p(w)=\frac{1}{\mathcal{S}}$ for all $w\in\mathbb{S}$ where $\mathcal{S}=|\mathbb{S}|$ is the codebook cardinality, so that Eq.~\eqref{eq:exclusion_strategy_generic} can be rewritten as a stochastic mapping
\begin{equation}
    \tilde v_{k\subscript{NULL}}(\hat w\vert \hat s)=\begin{cases}
        \frac{1}{\mathcal{S}-1}, & \hat w\in\mathbb{S}\backslash\{\hat s\}\\
        0, & \text{otherwise}
        \end{cases}.
        \label{eq:pdf_exclusion_strategy}
	\end{equation}
\end{enumerate}
We let the receiver to opportunistically select one of the three options by chance of $\beta_1$, $\beta_2$, and $\beta_3$, respectively, so that instead of the deterministic decryptor \eqref{eq:pdf_decryptor} we have now a stochastic one
\begin{equation}
	\tilde v_{\hat k}(\hat w\vert \hat s)=\begin{cases}
		1, & \hat w=f^{-1}_{\hat k}(\hat s), \hat k\neq k\subscript{NULL}\\
		\beta_1, & \hat w=\hat s, \hat k=k\subscript{NULL}\\
		\beta_2, & \hat w=s\subscript{NULL}, \hat k=k\subscript{NULL}\\
		\frac{\beta_3}{\mathcal{S}-1}, & \hat w\in\mathbb{S}\backslash\{\hat s\}, \hat k=k\subscript{NULL}\\
		0, & \text{otherwise}
	\end{cases},\label{eq:pdf_opportunistic_decryptor}
\end{equation}
leading to the distortion in Eq.~\eqref{eq:distortion_opportunistic_decryptor}, which is dependent on the packet transmission error rates over the primary and secondary transport channels. We denote it as $\tilde D\left(\varepsilon\superscript{p},\varepsilon\superscript{s}\right)$.

For \emph{Bob}, given certain channel conditions and radio resource allocation, $\varepsilon\superscript{p}$ and $\varepsilon\superscript{s}$ are determined, which we denote as $\varepsilon\superscript{p}\subscript{Bob}$ and $\varepsilon\superscript{s}\subscript{Bob}$, respectively. He can then select the optimal strategy $(\beta_1\superscript{Bob}, \beta_2\superscript{Bob}, \beta_3\superscript{Bob})$ to miminize the expected distortion $\tilde D$ by solving the optimization problem:
\setcounter{equation}{\theequation+1}
\begin{mini!}
	{\beta_1, \beta_2, \beta_3}{\tilde D\subscript{Bob}=\tilde D\left(\varepsilon\superscript{p}\subscript{Bob}, \varepsilon\superscript{s}\subscript{Bob}\right)}{\label{eq:optimization_problem}}{(OP):}{}
	\addConstraint{\beta_1+\beta_2+\beta_3}{=1},
\end{mini!}
which can be simply solved by comparing the terms $\Delta_1$, $\Delta_2$, and $\Delta_3$ in Eq.~\eqref{eq:distortion_opportunistic_decryptor}, which are associated with $\beta_1$, $\beta_2$, and $\beta_3$, respectively. For $i\in\{1,2,3\}$, the optimum always satisfy
\begin{align}
	\beta_i &= 0, \qquad \forall \Delta_i\neq\Delta\subscript{max}\\
	\sum\limits_{i:\Delta_i=\Delta\subscript{max}}\beta_i&=1,
\end{align}
where $\Delta\subscript{max}=\max\{\Delta_1,\Delta_2,\Delta_3\}$.

Similarly, \emph{Eve} can also select the optimal strategy $(\beta_1\superscript{Eve}, \beta_2\superscript{Eve}, \beta_3\superscript{Eve})$ to minimize its expected distortion $\tilde D\subscript{Eve}=\tilde D\left(\varepsilon\superscript{p}\subscript{Eve},\varepsilon\superscript{s}\subscript{Eve}\right)$.

\section{Deception Strategy Optimization}\label{sec:tx_strategy_optimization}
Now, consider \emph{Alice} who is aware of the opportunistic decryptor strategy of \emph{Bob} and \emph{Eve}. Given fixed \ac{mcs} and radio resource allocation, the packet error rates of \emph{Bob}, i.e., $\varepsilon\superscript{p}\subscript{Bob}$ and $\varepsilon\superscript{s}\subscript{Bob}$, can be easily estimated by \emph{Alice} based on her channel measurements. Therefore, \emph{Bob}'s optimal strategy of opportunistic decryption and the correspondingly minimized mean distortion, i.e., $\min(\tilde D\subscript{Bob})$, can be well predicted by \emph{Alice} as functions of her deceptive encryptor's activation rate $\alpha$. On the otherhand, though \emph{Alice} cannot directly measure the channel conditions of \emph{Eve}, it is common to consider her possessing the statistical knowledge about the eavesdropping channel, and therewith capable of estimating \emph{Eve}'s expected error probabilities, i.e. $\mathbb{E}\{\varepsilon\superscript{p}\subscript{Eve}\}$ and $\mathbb{E}\{\varepsilon\superscript{s}\subscript{Eve}\}$, respectively. Therewith, \emph{Alice} can also predict \emph{Eve}'s optimal strategy of opportunistic decryption and the correspondingly minimized mean distortion under such expected channel conditions, which we denote as $\min(\breve D\subscript{Eve})=\min\left(D\subscript{Eve}\left(\mathbb{E}\{\varepsilon\superscript{p}\subscript{Eve}\}, \mathbb{E}\{\varepsilon\superscript{s}\subscript{Eve}\}\right)\right)$, as functions of $\alpha$.

Such insight implies the possibility of \emph{Alice} to optimize her semantic secrecy performance by adjusting its deceptive encryptor's activation rate $\alpha$ upon the measured/estimated channel conditions, without respecifying the \ac{mcs} or power allocation. One example optimization problem of such kind can be formulated as:
\begin{maxi!}
	{\alpha}{\min\breve D\subscript{Eve}}{\label{eq:optimization_problem_alice}}{(OP 2):}{}
	\addConstraint{\min\tilde D\subscript{Bob}}{\leqslant D\subscript{max}},
\end{maxi!}
where $D\subscript{max}$ is a predefined threshold, which shall be carefully selected to ensure the feasibility of the problem.

\section{Numercial Results}\label{sec:numerical_results}
\begin{figure}[!htbp]
	\centering
	\begin{subfigure}[t]{.9\linewidth}
		\centering
		\includegraphics[width=0.9\linewidth]{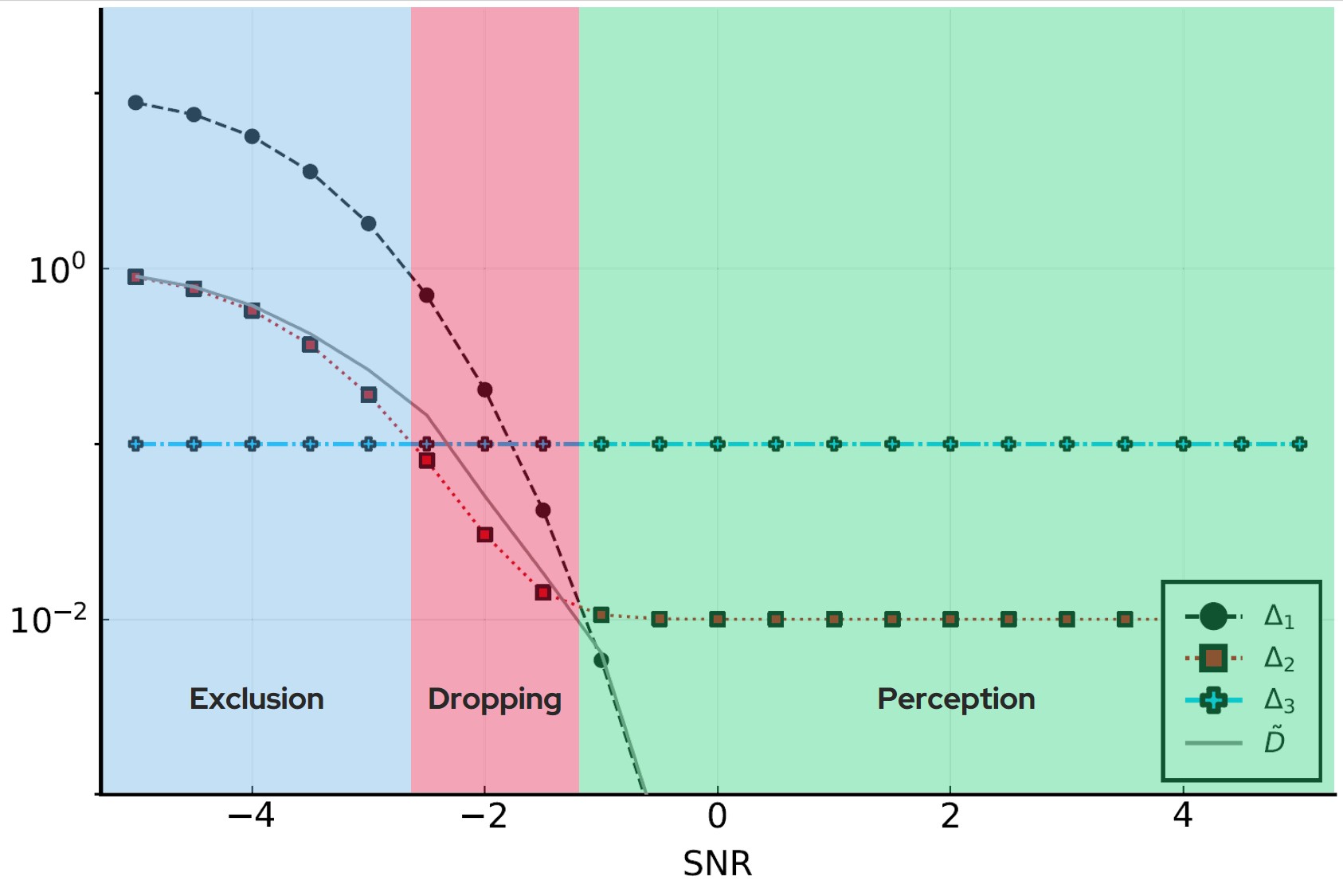}
		\subcaption{With a small codebook (cardinality $\mathcal{S}=2$)}
		\label{subfig:rx_low_card}
	\end{subfigure}
	\begin{subfigure}[t]{.9\linewidth}
		\centering
		\includegraphics[width=0.9\linewidth]{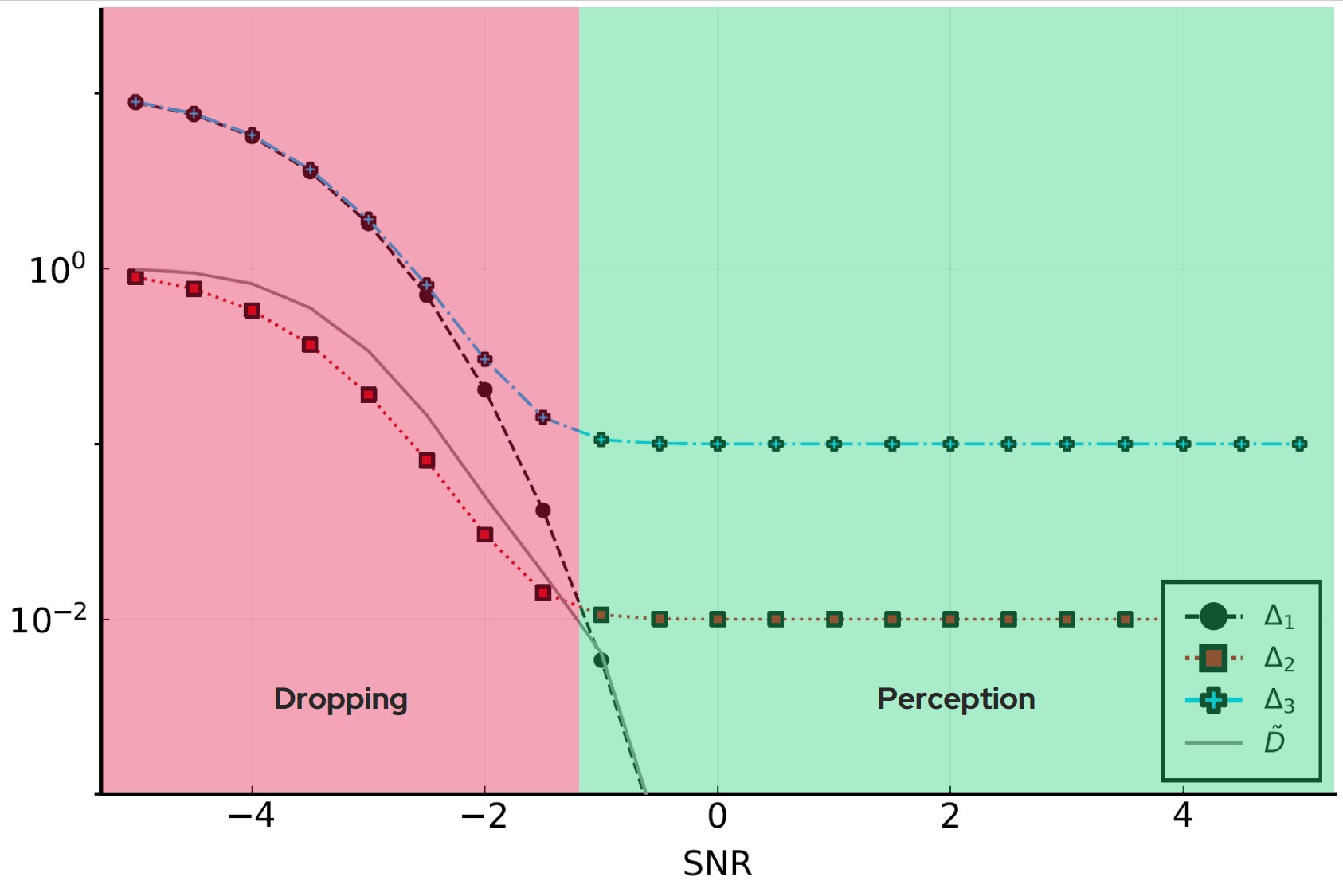}
		\subcaption{With a large codebook (cardinality $\mathcal{S}=2^{64}$)}
		\label{subfig:rx_high_card}
	\end{subfigure}
	\caption{The minimum semantic distortion and the strategy}
	\label{fig:rx_strategy}
\end{figure}
\begin{figure}[!htbp]
	\centering
	\begin{subfigure}[t]{\linewidth}
		\centering
		\vspace{-4mm}
		\includegraphics[width=0.85\linewidth]{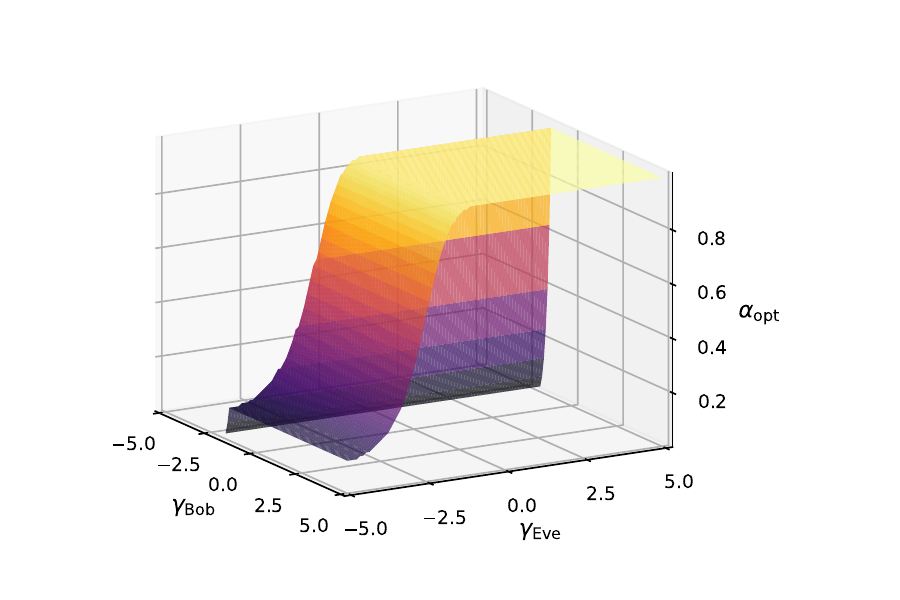}
		\subcaption{Optimal $\alpha$}
		\label{subfig:opt_alpha}
	\end{subfigure}\\
	\begin{subfigure}[t]{\linewidth}
		\centering
		\vspace{-0.5mm}
		\includegraphics[width=0.85\linewidth]{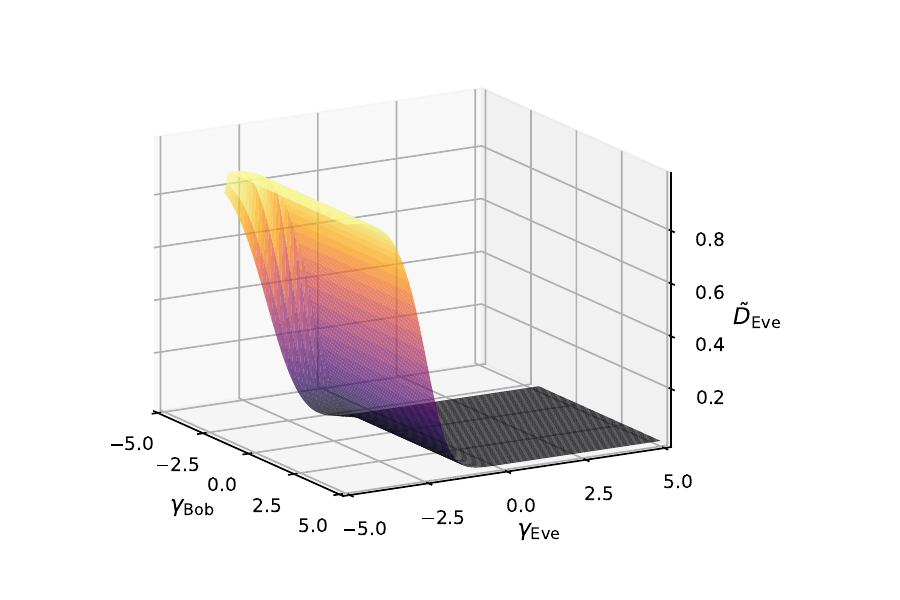}
		\subcaption{Eve's distortion}
		\label{subfig:eve_dist}
	\end{subfigure}
	\caption{Optimal deception strategy and performance}
	\label{fig:tx_strategy}
\end{figure}

\subsection{Opportunistic Receiving Strategy}\label{subsec:rx_strategy}
A case study on the opportunistic receiving strategy was carried out with numerical simulations. We set $D\subscript{loss}=1$ and $D\subscript{conf}=10$, with $\alpha=0.99$. 
Assuming sufficient channel coding redundancy, the erasure probabilities over the transport channels can be approximated with the packet loss rates~\cite{CHZ+2024physical}, which
% The packet loss rates over the primary and secondary transport channels 
were simulated regarding the \ac{fbl} theory~\cite{PPV2010channel},
where both $s$ and $k$ were set with the payload length of 64 bits and channel encoded with coding rate of $1/2$. The radio channel $T$ was, following the common routine in the field of \ac{fbl} information theory, normalized to unit bandwidth and evaluated over the range between \SI{-5}{\dB} and \SI{5}{\dB}. We repeated the study twice with different cardinalities $\mathcal{S}$ of the valid codebook $\mathbb{S}$, once miminized to $2$ and once maximized to $2^{64}$. The results are depicted in Fig.~\ref{fig:rx_strategy}. As expected, we observe that the \emph{perception} strategy is optimal at high \ac{snr}, while the \emph{dropping} strategy is preferred when the \ac{snr} drops into lower ranges. With a sufficiently poor \ac{snr}, the \emph{exclusion} strategy may come into play, but only when the codebook is sufficiently small.

\subsection{Optimal Deception Strategy}
Then, we numerically investigated the optimal deception strategy by \emph{Alice} with the same system specifications as in Sec.~\ref{subsec:rx_strategy}, taking the codebook cardinality as $2^{64}$ and $D\subscript{max}=0.01$. The results are shown in Fig.~\ref{fig:tx_strategy}. We observe that the optimal deception strategy $\alpha$ motonotically increases with \emph{Eve}'s \ac{snr} $\gamma\subscript{Eve}$, while \emph{Eve}'s distortion $\breve D\subscript{Eve}$ decreases with $\gamma\subscript{Eve}$. The dependencies on \emph{Bob}'s \ac{snr} $\gamma\subscript{Bob}$ is only observable at the boundary of the feasible region, due to the fixed $D\subscript{max}$. Nevertheless, it shall be noted that when a flexible adjustment of the radio resource allocation is allowed, the achievable system performance will be highly dependent on \emph{Bob}'s channel condition, similar to the results presented in our preliminary works~\cite{HZS+2023non,CHZ+2024physical}.

\section{Conclusion and Outlooks}\label{sec:conclusion}
% In this paper, we have proposed a novel semantic communication model for \ac{pld} systems, which provides a new perspective to understand the \ac{pld} mechanism and new insights to it. We have analyzed the semantic distortion in \ac{pld} systems with different levels of receiver's knowledge about the encryption key, and proposed an opportunistic receiving strategy for the receiver to minimize its semantic distortion. We have also investigated the optimal deception strategy for the transmitter to maximize the semantic secrecy performance. Numerical results have shown the potential of a cost-efficient adaptive \ac{pld} approach by means of adjusting the deception rate upon channel conditions.
In this paper, we have proposed a novel semantic communication model for \ac{pld}, which provides a new perspective to understand this mechanism and new insights to it. We have analyzed the semantic distortion in \ac{pld} systems with different levels of receiver's knowledge, and proposed an opportunistic receiving strategy to minimize the distortion. We have also investigated the optimal deception strategy to maximize the semantic secrecy performance. Numerical results have shown the potential of a cost-efficient adaptive \ac{pld} approach by means of adjusting the deception rate upon channel conditions.

As a planned future extension of this work, the generality of our proposed model can be further improved by analyzing the accurate erasure probabilities, instead of taking the packet error rates as their approximates under certain assumptions.

% \section*{Acknowledgment}
% This work is supported by the German Federal Ministry of Education and Research in the programme of ``Souver\"an. Digital. Vernetzt.'' joint projects 6G-RIC (16KISK028), Open6GHub (16KISK003K/16KISK004/16KISK012), and 6G-ANNA(16KISK105). B. Han (bin.han@rptu.de) and Y. Zhu (yao.zhu@inda.rwth-aachen.de) are the corresponding authors. 

\bibliographystyle{IEEEtran}
\bibliography{references}

\end{document}